\documentclass[10pt,amsmath, amssymb, aps,prb]{revtex4-2}
\usepackage{amsmath,amssymb}
\usepackage[dvips]{graphicx} 
\usepackage{dcolumn} 
\usepackage{bm} 
\usepackage{xcolor} 
\usepackage{feynmp-auto}
\usepackage{comment}

\usepackage{hyperref}
\usepackage[mathlines]{lineno}
\usepackage{epsfig,lineno}

\usepackage{scalerel}
\usepackage{tikz}
\usetikzlibrary{svg.path}

\definecolor{orcidlogocol}{HTML}{A6CE39}
\tikzset{
  orcidlogo/.pic={
    \fill[orcidlogocol] svg{M256,128c0,70.7-57.3,128-128,128C57.3,256,0,198.7,0,128C0,57.3,57.3,0,128,0C198.7,0,256,57.3,256,128z};
    \fill[white] svg{M86.3,186.2H70.9V79.1h15.4v48.4V186.2z}
                 svg{M108.9,79.1h41.6c39.6,0,57,28.3,57,53.6c0,27.5-21.5,53.6-56.8,53.6h-41.8V79.1z M124.3,172.4h24.5c34.9,0,42.9-26.5,42.9-39.7c0-21.5-13.7-39.7-43.7-39.7h-23.7V172.4z}
                 svg{M88.7,56.8c0,5.5-4.5,10.1-10.1,10.1c-5.6,0-10.1-4.6-10.1-10.1c0-5.6,4.5-10.1,10.1-10.1C84.2,46.7,88.7,51.3,88.7,56.8z};
  }
}

\newcommand\orcidicon[1]{\href{https://orcid.org/#1}{\mbox{\scalerel*{
\begin{tikzpicture}[yscale=-1,transform shape]
\pic{orcidlogo};
\end{tikzpicture}
}{|}}}}

\usepackage{hyperref}

\usepackage{graphicx}
\usepackage{dcolumn}
\usepackage{bm}
\usepackage{subfigure}
\usepackage{epstopdf}
\usepackage{hyperref}
\usepackage{color}
\usepackage[utf8]{inputenc}

\newcommand{\gen}{GEn}
\newcommand{\gep}{GEp}
\newcommand{\gmn}{GMn}
\newcommand{\gmp}{GMp}
\newcommand{\ALL}{$A_{_{LL}}$}
\newcommand{\KLL}{$K_{_{LL}}$}

\newcommand{\gevsq}{(GeV/$c$)$^2$}
\newcommand{\qsq}{Q$^2$}

\begin{document}
\title{\bf {The Super Bigbite Spectrometer physics program}}
\author{B.~Wojtsekhowski \orcidicon{0000-0002-2160-9814}} 
\email[Contact person, ]{bogdanw@jlab.org} 
\affiliation{Thomas Jefferson National Accelerator Facility, Newport News, VA 23606}
\author{G.~Cates \orcidicon {0000-0002-9373-9271}}
\affiliation{University of Virginia, Charlottesville, VA 23606}

\begin{abstract}
{The structure of the nucleon is a central problem in strong interaction physics in the non-perturbative regime. 
Indeed, the vast majority of the known matter in the Universe is made of protons and neutrons which are a remarkable emergent phenomenon of quantum chromodynamics. 
A critical aspect of investigating nucleon structure experimentally is the measurement of fundamental quantities such the elastic nucleon form factors.  Also important is the measurement transverse momentum dependent distribution functions.  Accessing such quantities experimentally, however, is challenging because of the small cross sections involved, particularly at high momentum transfer. We present here a physics program that is addressing this challenge based on the Super Bigbite Spectrometer (SBS) that has recently been built at the Thomas Jefferson National Accelerator Facility.  SBS provides a relatively large solid angle of 70 msr and can be used at high luminosities and forward-scattering angles. It is based on a single large dipole magnet in an open-geometry in which the detector package has a direct line of sight to the target.  This approach is only possible through the use of detector technology that can operate at very high rates while providing excellent spatial resolution.  It is the product of solid angle and luminosity that is critical when measuring small cross sections, and in this regard, among spectrometer systems at JLab, SBS is presently unique in its capability.  The first set of experiments utilizing SBS has been successfully completed, and more experiments are planned for the future.  We also discuss a proposed upgrade that would increase the SBS solid angle to 260 msr, thereby opening perspectives for an even broader physic program.}
\end{abstract}
\date{\today}
\maketitle
%
\large

\section{Introduction}

The study of the structure of hadronic matter using electron scattering has been very productive and has led to multiple discoveries related to the nucleon including the finite size of a proton~\cite{Hofstadter}, scaling in the deep inelastic regime~\cite{DIS}, and nucleon spin structure~\cite{EMC}, as well as the recent quark  flavor decomposition of the ground-state elastic form factors~\cite{Flavor}. 
With the discovery of asymptotic freedom~\cite{qcd}, quantum chromodynamics (QCD) has emerged as the correct theory of strong interactions, although rigorous {\it ab initio} calculations have been largely confined to the perturbative regime, and calculations of the static properties and dynamic structure functions are limited to computational methods such as lattice QCD.  Historically, a description of the nucleon has relied on measurements of the elastic and transition form factors as well as Parton Distribution Functions (PDFs) that are measured in deep inelastic scattering. A huge step forward has been the development of Generalized Parton Distributions (GPDs) that provides a unified QCD-based theory within which PDFs and form factors can all be viewed as being derived from Wigner distributions~\cite{GPD-1,GPD-2,GPD-3}. 
Comparisons between experimental results and various advanced calculations, including those based on the Dyson-Schwinger Equations~\cite{DSE}, the lattice QCD~\cite{lQCD-FF}, as well as those that employ GPD-based models~\cite{DK-2013}, have provided deeper physical understanding. 
Such comparisons become more accurate at high momentum transfer ($Q^2$), but face the challenge that the relevant cross sections become quite small at high $Q^2$; elastic scattering cross sections, for example, decrease roughly like $Q^{-12}$.

By the early 1980s, the US nuclear physics community formulated a wide program of investigations which required a new electron accelerator facility.
It was approved for construction by the Department of Energy in 1983, see refs.~\cite{CEBAF-1985, CEBAF-1987}. 
The CEBAF electron accelerator~\cite{CEBAF-2001, CEBAF-2024} has been providing a high-intensity polarized continuous wave beam for many experiments since 1995. Initially CEBAF had a maximum energy of 6~GeV but was later upgraded to 12 GeV.

The Super Bigbite Spectrometer (SBS) was initially developed with the elastic nucleon form factors and their small cross section at high $Q^2$ in mind~\cite{BW2014}.
The SBS program has subsequently expanded to include, for example, the exploration of up- and down-quark Transverse Momentum Distributions (TMDs) at a scale of 1-2\% of the nucleon's size.  It is important to note that at present, the elastic nucleon form factors provide one of the strongest constraints that exist on GPD models. Results from SBS will also be used for the determination of the up- and down-quark contributions to the Pauli and Dirac form factors and will clarify the evidence for di-quark correlations~\cite{ECT-2019,Q+DQ-2025}. The study of $F_{1(2)}^{u(d)}$~\cite{Flavor} provided direct evidence of di-quark correlations in the nucleon, and extending those observations to higher momentum transfer is of enormous importance. 

The SBS is based on a single dipole magnet and allows a x10 increase in solid angle and momentum acceptance compared to typical high-resolution spectrometers.
After the approval of the original proposals~\cite{GEp,GEn-2,GMn-1} on nucleon form factors, the physics program has widened and now includes a study of the pion PDF using a tagged-DIS (Sullivan) process~\cite{TDIS}, polarized TMDs in semi-inclusive DIS~\cite{SIDIS}, the neutron electric form factor via the polarization transfer method~\cite{GEn-RP}, a precision measurement of the neutron's electric form factor using a Rosenbluth separation (to help isolate two-photon effects)~\cite{nTPE}, double polarization asymmetries in wide-angle pion photo-production~\cite{high-s-Pion}, wide angle Compton scattering from a proton at large s, |t|~\cite{WACS-ALL}, and two experiments with a positron beam~\cite{GEp+,nTPE+}.
Recently, two more proposals were developed: $\phi$-meson electro-production at large momentum transfer~\cite{deepPhi} and neutrino electro-production~\cite{tAVFF}.
Assuming plans move forward to develop a version of SBS with significantly larger solid angle, it also appears possible to make a high accuracy measurement of parity violation in DIS at high $Q^2$.

\section{Instrumentation}

Achieving good event rates can be quite challenging in electron-scattering experiments involving high momentum transfer. Luminosity, the product of the electron-beam intensity and the areal density of the target, needs to be as large as possible. Also critical is the solid angle of the detection system.  The productivity of the experiment relies on both of these factors, which must be optimized, subject to constraints such as the detector rate capability. With the advancement of tracking technology, however, such as the utilization of gas electron multiplier (GEM) technology~\cite{Sauli} and highly segmented fast multi-wire drift chambers (MWDC)~\cite{MWDC} as well as streaming data processing, restrictions from detector counting rates are significantly relaxed.

\begin{figure}[htb]
\begin{center}
\includegraphics[trim = -20mm 0mm 0mm 0mm, angle=0, width=0.75\textwidth]{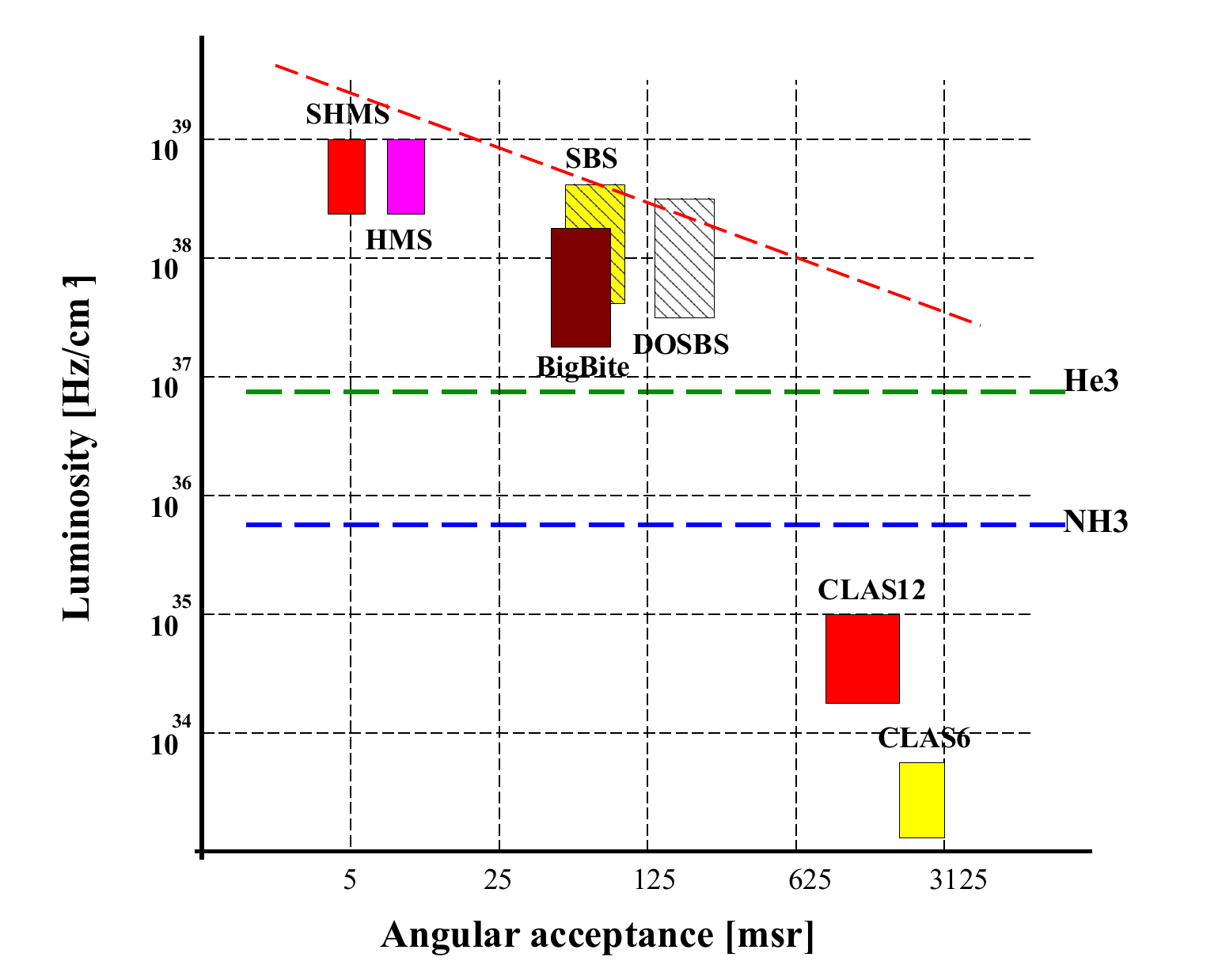}
\caption{Usable luminosity vs. solid angle of the JLab spectrometers. 
The green dashed line shows the limit for a polarized $^3$He~target. 
The blue dashed line shows the limit for a polarized NH3 target.
The red dashed line shows the achievable product of the luminosity and the detector solid angle in an experiment.}
\label{fig:Landscape}
\end{center}
\end{figure}

Fig.~\ref{fig:Landscape} presents the landscape of the JLab detector systems with usable luminosity shown on the vertical axis and solid angle acceptance shown on the horizontal axis. The dot-dashed red line represents values of luminosity and solid angle for which their product is constant and, all other things remaining equal, for which the  event rates would be constant. We have drawn the line through the region covered by SBS to facilitate comparison with JLab's other detector systems.
The SHMS and HMS spectrometers in Hall~C provide high momentum resolution and are capable of operating at a very high luminosity of $10^{39}$~cm$^{-2}\rm/s$ but have a relatively small solid-angle acceptance. 
The large acceptance detector in Hall~B, CLAS (6 GeV era) and CLAS12 (12 GeV era), has a solid angle around a few~sr, but maximum luminosity is limited to roughly $10^{35}$~cm$^{-2}/\rm s$.
SBS and BB have middle-sized solid angle of 70-100 msr and usable luminosity up to $3\times10^{38}$~cm$^{-2}\rm/s$.
Also shown on Figure~\ref{fig:Landscape} with the horizontal dashed red and blue lines are estimates of the maximum luminosities (at the time of this writing) that are practical with gaseous polarized $^3$He and solid polarized NH$_3$ targets.
Figure~\ref{fig:Landscape} shows the advantage of the SBS/BB system especially in experiments with a polarized $^3$He target.

\subsection{Super Bigbite Spectrometer}
\label{sec:sbs}

In elastic scattering at high \qsq, the event selection is complicated by the large counting rate from inelastic processes. An effective strategy for isolating the low cross-section elastic events is to detect  both final-state particles (the electron and the nucleon) in coincidence.
An important issue in such measurements is the angle of the recoil nucleon relative to the beam direction. For a given electron energy, the \qsq\  goes up as the electron's scattering angle is increased, but the angle of the recoil becomes progressively smaller. Designing a spectrometer for use at small scattering angles usually leads to a big loss of solid angle due to the need to increase the distance between the target and the spectrometer to accommodate  the width of the spectrometer magnets. 
\begin{figure}[!htb]
\begin{center}
\includegraphics[trim = 80mm 30mm 50mm 30mm, angle=0, width=0.5\textwidth]{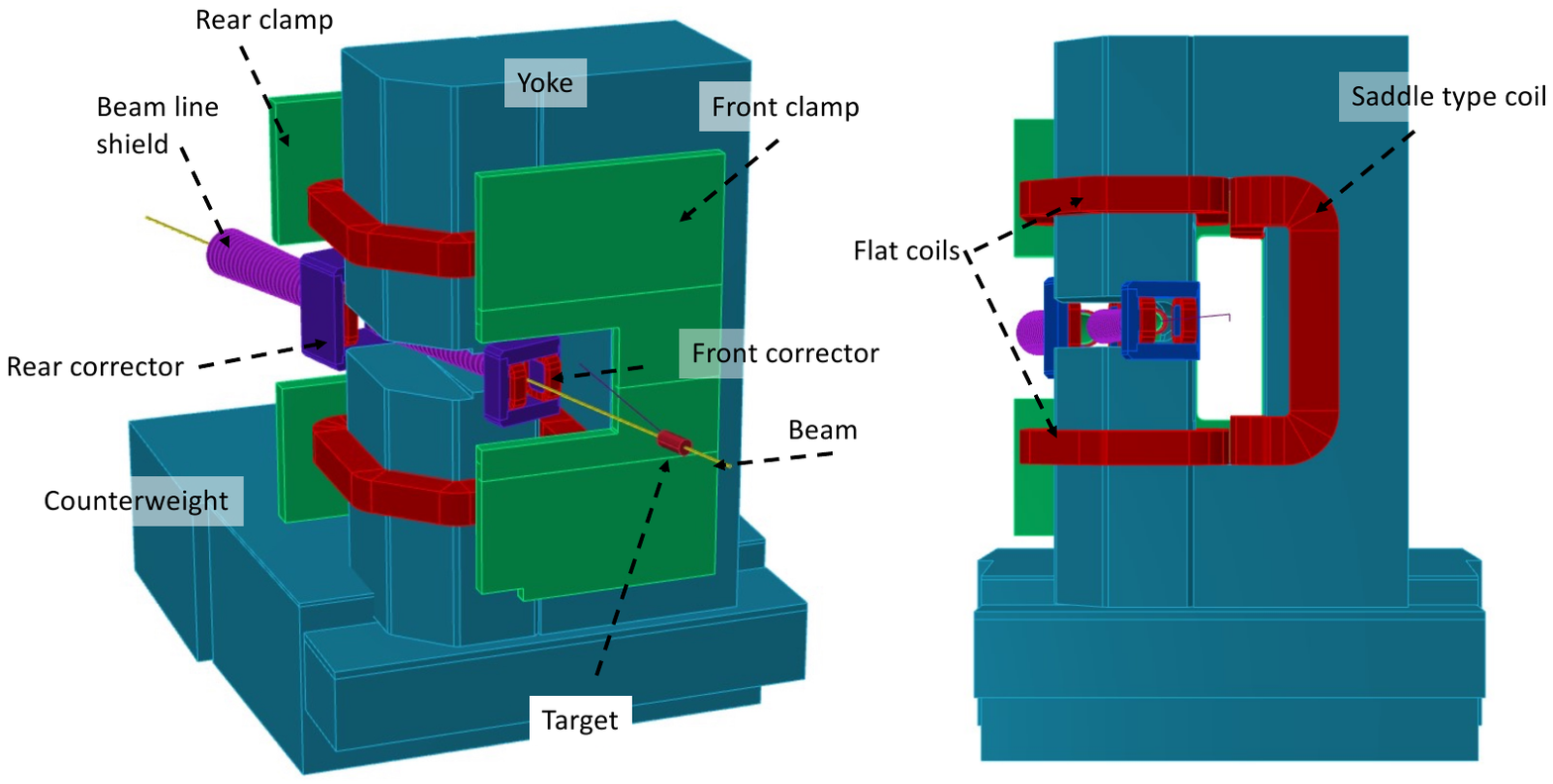}
\caption{The views of the SBS magnet. 
A 3D view on the left and a front view on the right (a front field clamp removed for visibility of structure).}
\label{fig:3D-view}
\end{center}
\end{figure}

\begin{figure}[!ht]
\begin{center}
\includegraphics[trim = 30mm 30mm 30mm 30mm, angle=0, width=0.75\textwidth]{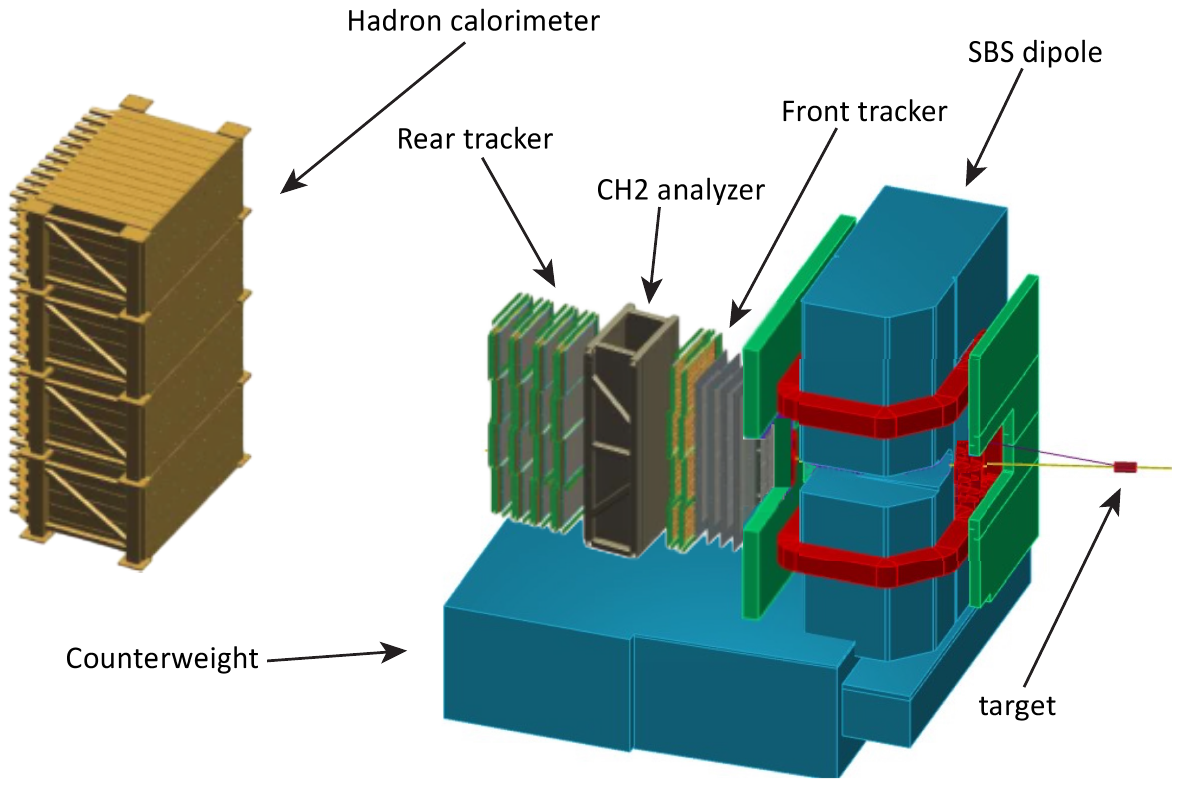}
\caption{The view of the SBS with the detector package of the GEp experiment.}
\label{fig:SBS-GEp}
\end{center}
\end{figure}

We have found a solution to maintaining a relatively large solid angle even when placing our spectrometer at fairly small angles with respect to the beam line.  
The key is making a narrow cut in the spectrometer yoke to accommodate the beam line downstream of the target as can be seen in Fig.~\ref{fig:3D-view}. 
The effect on both the strength and homogeneity of the magnetic field in the spectrometer aperture is relatively minor and does not appreciably affect the moderate momentum resolution that is inherent in our single-dipole open-geometry detector design.
The resulting solid angle is 70~msr for scattering angles above 15~degrees and, as illustrated in Fig.~\ref{fig:Landscape}, ensures that for an important class of experiments, the SBS has exceptional performance when compared with other detector systems at JLab.
A detailed description of the SBS magnetic concept is presented in ref.~\cite{SBS-spectrometer}.
An essential feature of the SBS spectrometer is the absence of a shielding hut, which is
possible due to the high energy threshold of the trigger.
As a result, the detector package and location of its components are flexible, e.g. in the GMn experiment~\cite{GMn-1} the hadron calorimeter (a trigger) was located at a distance of 17~m from the target.
Figure~\ref{fig:SBS-GEp} shows the SBS with the detector package used in the GEp experiment that measured the ratio of the electric and magnetic form factors of the proton (see Section \ref{sec:gep} and ref.~\cite{GEp-update}).

\subsection{Big Bite spectrometer}
\label{sec:bigbite}
\begin{figure}[ht]
\begin{center}
        \includegraphics[trim = 0mm 30mm 0mm 30mm, angle=0, width=0.9\textwidth]{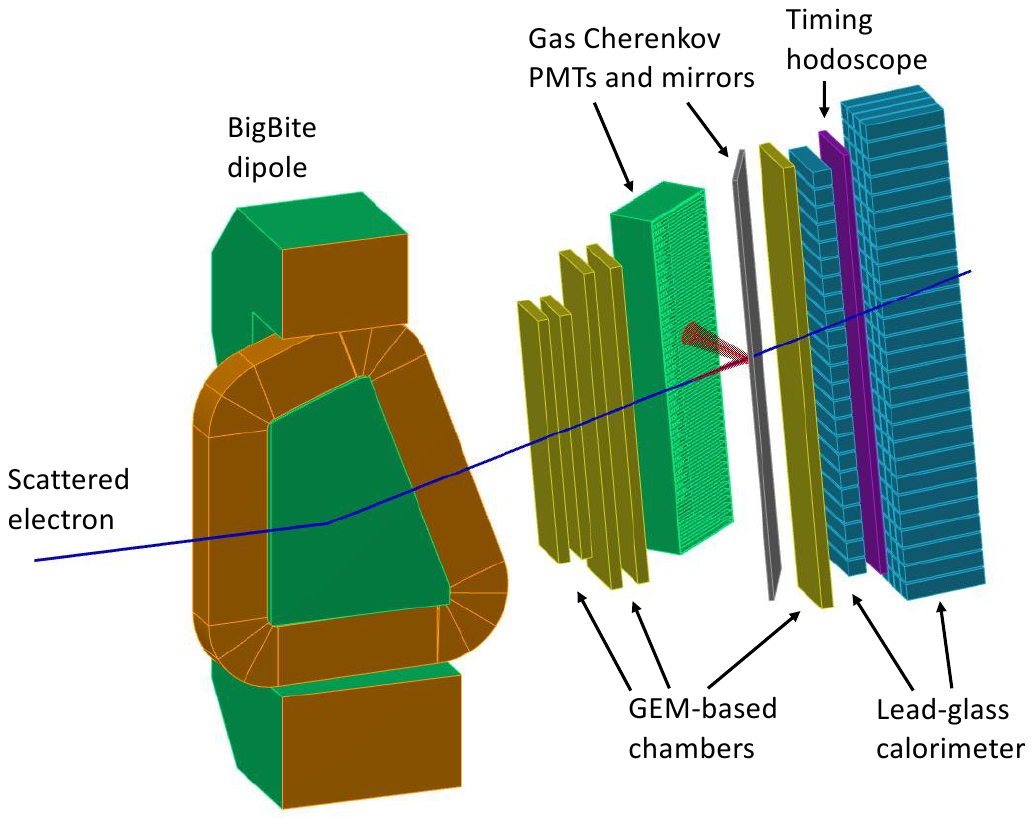}
    \caption{Cutaway view of the BBS magnet and the detector package.} 
    \label{fig:BB}
\end{center}
\end{figure}

The electron arm for high \qsq~measurements requires a spectrometer at a central angle of 30-50 degrees.
The BigBite spectrometer based on a NIKHEF-K magnet~\cite{Doug} provides significant advantages due to its large solid angle and momentum bite and has been used for several experiments in the SBS program~\cite{GEn-1,Pion,SIDIS-1}.
The detector package of BigBite includes a new tracker with five GEM-based chambers~\cite{GEM}, a highly segmented gas Cherenkov counter, a high resolution timing hodoscope, and a two-layer electromagnetic calorimeter~\cite{BBCal} as shown in Fig.~\ref{fig:BB}. 

\subsection{Large area GEM-based tracker}
A GEM-based coordinate detector~\cite{Sauli} has also been constructed for SBS~\cite{GEM}.
For the GEp experiment~\cite{GEp}, two large trackers were made.
Each tracking detector includes 8~GEM-based planes.
The individual plane dimensions vary: 40~cm x 150~cm in the six front planes and 60~cm x 200~cm in the remaining 10~planes. 
Coordinate resolution of 70~$\mu$m in each plane was obtained.

\subsection{Hadron detector (HCAL)}
\label{sec:hcal}
Another key component of the hadron arm detector is a large (1.8~m by 3.6~m) highly segmented (12x24) 
calorimeter which allows a high energy threshold for the DAQ trigger with good time resolution of 0.75~ns (rms) and coordinate resolution of 5.5~cm.
For SBS experiments where tracking is used in the hadron arm (such as GEp~\cite{GEp-update}), it provides a starting coordinate for the track search which is critical due to the large occupancy in the tracker at the required operating luminosity.

\subsection{Polarized He-3 target}

The primary SBS experiment that probed the electric form factor of the neutron, GEn-II, utilized a polarized $^3$He target \cite{GEn-2}.   
The isotope $^3$He nucleus makes an excellent polarized neutron target because the sole neutron carries most of the spin (1/2) while the protons are mostly paired to spin zero \cite{Woloshyn}. 
Furthermore, gaseous polarized $^3$He targets can tolerate relatively high luminosities, particularly compared to polarized solid targets used for spin-dependent studies of the proton and deuteron. 
Prior to SBS, the utility of polarized $^3$He targets for measuring \gen~has been demonstrated in several experiments~\cite{MAMI-1, GEn-1, MAMI-2}.

The performance of polarized $^3$He targets has steadily improved since some of the first applications in electron scattering such as the measurement of the spin structure of the neutron by E142 experiment~\cite{E142} at Stanford Linear Accelerator Center (SLAC). 
The improvements have been due to both better understanding of the underlying physics as well as improvements in laser technology. 
The polarized $^3$He target used for GEn-II was based on the technique of spin-exchange optical pumping~\cite{Baranga}. 
It utilized a hybrid mixture of both Rb and K in the polarization process in contrast to early targets that used Rb alone. 
Alkali-hybrid mixtures greatly increase the efficiency with which the $^3$He is polarized, resulting in significantly higher performance targets as discussed by Singh {\it et al.}~\cite{Singh}. 
Polarized $^3$He targets based on spin-exchange optical pumping utilize high-pressure sealed glass cells, and the GEn-II target cells employed a novel mechanical design that made it possible to actively drive convection between the ``pumping chamber", in which the $^3$He is polarized, and the ``target chamber", through which the electron beam passes~\cite{High-t, PRC-He3}.  
GEn-II was only the second experiment to utilize ``convection-based" target cells and was run at a luminosity (electron-$\,^3$He) of roughly $\rm4.5\times10^{36}\,cm^{-2}\,s^{-1}$.  
When polarization is also taken into account, the figure-of-merit of the GEn-II target was more than a factor of 50 higher than that used during SLAC E142.

\subsection{Radiation hard electromagnetic calorimeter (ECAL)}
\label{sec:ecal}

A large area (3.3~m$^2$) highly segmented lead-glass electromagnetic calorimeter (ECAL) was built for the SBS GEp experiment~\cite{GEp}. Lead-glass calorimeters develop color centers and reduced transparency when exposed to even relatively small amounts of radiation (1 kRad). 
Common approaches to dealing with this problem include annealing the lead glass components through exposure to UV radiation, a procedure that is time consuming and results in down time if needed during an experiment.  
ECAL, however, employed a novel solution for maintaining transparency by maintaining the lead-glass blocks at high temperature ($\sim200$ C$^\circ$) throughout the experiment. This approach allowed us to operate at $3 \times 10^{38}$~cm$^{-2}/\rm s$ luminosity without noticeable loss of glass transparency over a four-month-long run.

\subsection{Compact Photon Source}
\label{sec:CPS}
We will describe in Section~\ref{sec:wacs} an SBS experiment that has been proposed to measure polarization observables in wide-angle Compton scattering. The experiment would utilize a solid polarized target such as NH$_3$~\cite{Crabb,Meyer}, which normally operates at a relatively low luminosity because of limitations due to beam heating. The targets typically operate at temperatures below $\rm1\,K$ and cannot tolerate  electron-beam currents above something like $\sim$100~nA.

\begin{figure}
\begin{center}
\includegraphics[trim = 90mm 0mm 80mm 0mm, angle=0, width=0.55\textwidth]{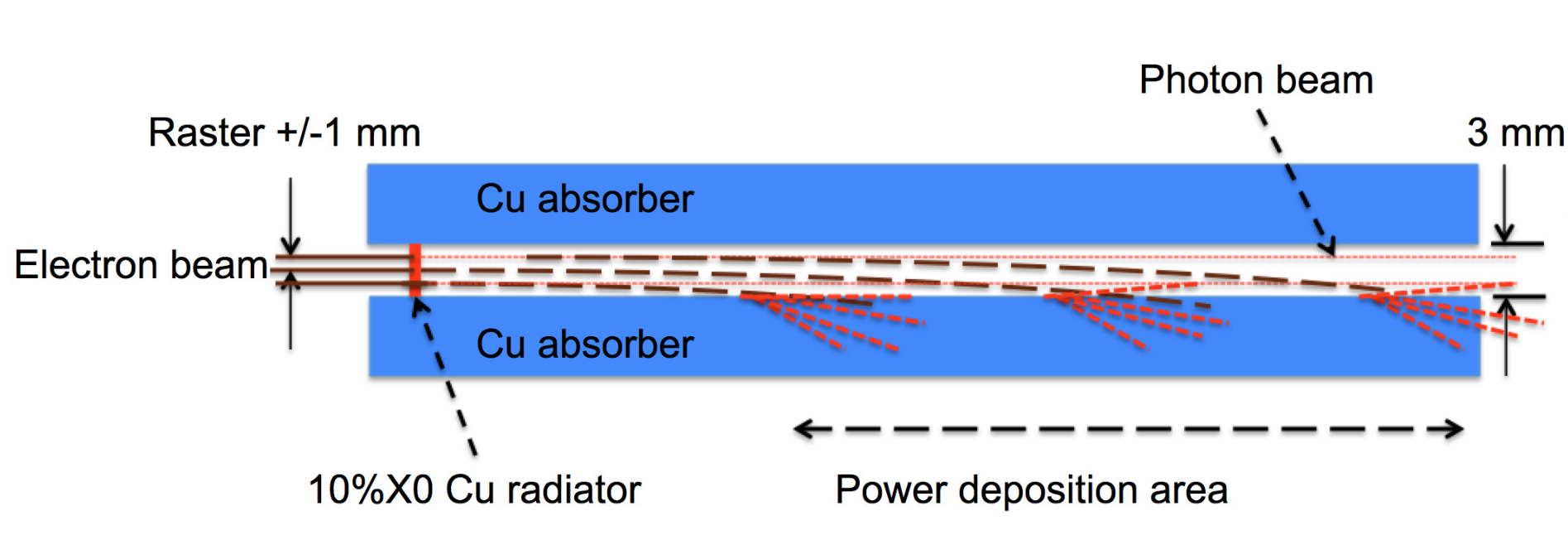}
\caption{\label{fig:CPS} The concept of the beam absorber in the Compact Photon Source.
 Combination of the narrow opening (sufficient for the outgoing photon beam), an electron beam raster, and small deflection angle, together allow wide distribution of the beam power along the absorber.
Figure is taken from ref.~\cite{CPS-2}.}
\end{center}
\end{figure}

Solid polarized targets can be operated at much higher luminosities if one uses a photon beam instead of an electron beam. The space requirements for photon beam preparation at 5-15 GeV~energy, however, are typically around 15-30~meters along the beam line~\cite{Cornell,SLAC-photo}.
For the above mentioned wide-angle Compton scattering experiment~\cite{WACS-ALL}, we have developed a novel design, Compact Photon Source, that only requires a few meters of space. 
It provides a photon beam intensity 30~times higher than would be the case for an equivalent heat load when using a configuration in which a mix of photons and electrons pass through the target~\cite{CPS-1,CPS-2}. 
Furthermore, our design delivers this flux while maintaining a small beam spot size at the target, which is useful for background suppression.
The main new component of the novel photon beam source is a beam deflecting magnet combined with a compact absorber for the electron-beam energy as described in ref.~\cite{CPS-2}. 
The central part of the system is illustrated in Fig.~\ref{fig:CPS}. 

\subsection{Double Open Super Bigbite Spectrometer}
\begin{figure}[!htb]
\begin{center}
\includegraphics[trim = 5mm 30mm 20mm 30mm, angle=0, width=0.9\textwidth]{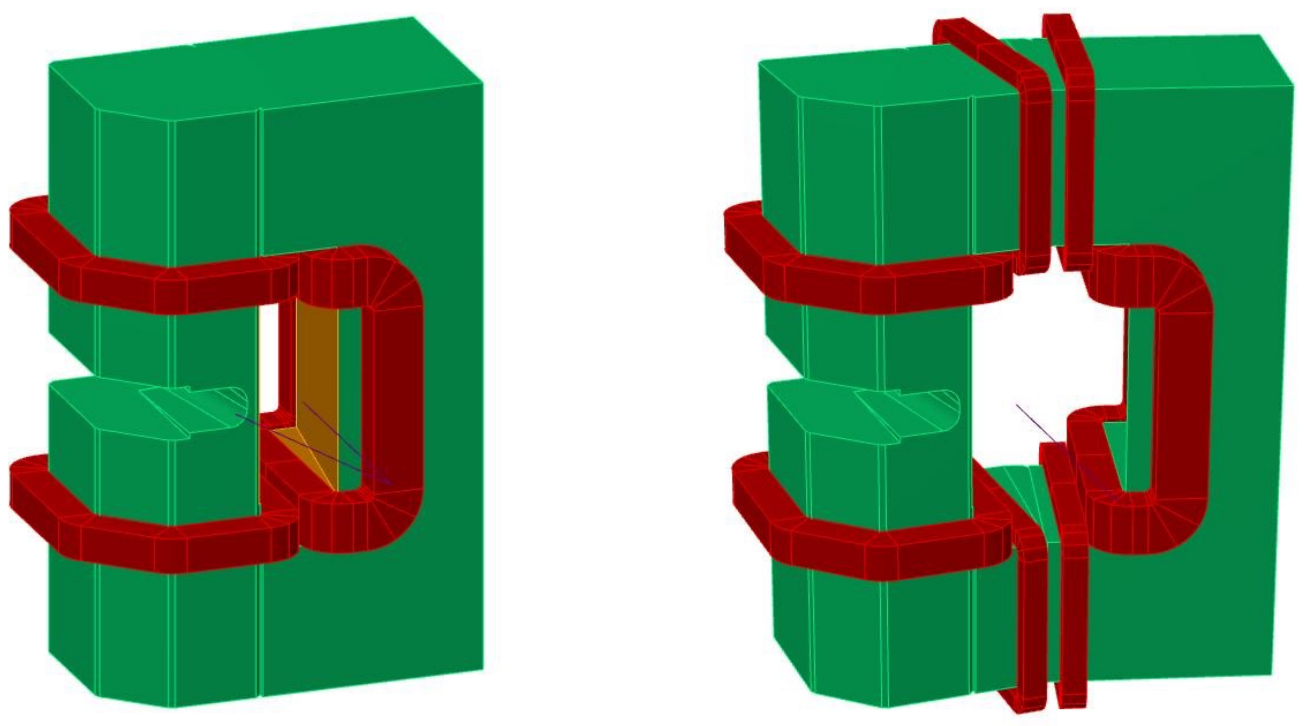}
\caption{The 3D view of the Open SBS (on the left) and Double Open SBS (on the right).
The highlighted surface inside the Open SBS shows the required modification of the yoke.}
\label{fig:DOSBS}
\end{center}
\end{figure}

We mention finally the possibility of increasing the solid angle of the SBS magnet by as much as a factor of four with relatively modest changes to the magnet geometry.
Such a development is motivated by three proposed experiments~\cite{deepPhi,tAVFF, PVDIS} that will be described in Section~\ref{sec:double}.
The concept of the modifications is illustrated in Fig.~\ref{fig:DOSBS}.

A particularly straightforward change is to reshape the gap between the two poles such that the downstream portion of the gap (the exit from the magnet) is wider than the upstream portion of the gap (the entrance to the magnet); such a change would result in a surprisingly large increase in the solid angle from 70~msr to 130~msr.

An even larger increase in the pole gap can be accomplished by inserting additional iron pieces between the two halves of the yoke.  
The SBS magnet and portions of its supporting structure were constructed from two ``48D48" magnets that were donated by Brookhaven National Laboratory. The additional iron pieces needed to substantially increase the pole gap already exist unused from the two 48D48 magnets and would need only minor modifications. The resulting spectrometer, that we refer to as the Double Open Super Bigbite Spectrometer (DOSBS), would reach 260~msr at a central angle of 28$^{\circ}$.

\section{High momentum transfer elastic nucleon form factors}

The study of the nucleon and its structure is a natural way of investigating the physics of QCD. Historically, a detailed understanding of atomic structure was critical to establishing the validity of Quantum Electrodynamics. Also, the study of the atomic nucleus has, over many decades, provided a detailed understanding of strong interactions from a perspective in which protons and neutrons are the primary degrees of freedom.  Understanding QCD, however, requires dissecting the nucleon in ways that unveil the roles of the quarks and gluons. This endeavor began with studies of deep inelastic scattering (DIS) but has necessarily grown in multiple directions.  The study of the elastic nucleon form factors plays an important role in this endeavor.  

The elastic nucleon form factors represent a set of observables that have a particularly well-defined theoretical interpretation.  Historically, elastic form factors provided the first direct measurements of the size of nucleons, and they indeed encode critical information about the spatial distribution of charge and magnetization.  Furthermore, the evolution of the elastic form factors at high values of  momentum transfer provides insight into the manner in which QCD governs nucleon structure. With the assumption of charge symmetry, elastic scattering data from the proton and the neutron can be combined to determine the contributions to the elastic form factors from the individual quark flavors. The elastic form factors also provide what are presently some of the tightest constraints on generalized parton distributions (GPDs) which promise to form the basis of what is essentially 3D imaging of nucleon structure.

If the goal is to resolve the underlying structure of the nucleon, it is important to probe the nucleon with the highest possible resolution. The length scale being probed in a scattering process goes like $(Q/\hbar)^{-1}$, a quantity that is equal to one fermi when $Q\sim 0.2\,\rm GeV/c$. The accelerator at Jefferson Laboratory provides an electron beam with energy up to 11~GeV which enables the study of scattering processes with $Q^2$ as high as 20~\gevsq, a length scale that corresponds to roughly 0.04 fermi.  The cross section for elastic scattering from protons and neutrons, however, drops off quickly with increasing momentum transfer, roughly like $Q^{-12}$. This precipitous drop with $Q^2$ reflects the fact that the probability of the nucleon holding together in its original form rather than undergoing some inelastic process becomes quite small at high energy scales. The design of the SBS focused on a system capable of providing good statistics for the low-probability elastic events with enough specificity to distinguish the elastic events from the enormous inelastic background. For the electric form factors of the proton and neutron, SBS was also designed to use the double-polarization method for reasons we discuss next.

Historically, the least well known of the elastic nucleon form factors have been the electric form factors of the proton and neutron, $G_E^p(Q^2)$ and $G_E^n(Q^2)$, respectively. It is straightforward to understand why this is the case.
In the one-photon-exchange approximation the elastic electron-nucleon  differential scattering cross section, $d\sigma (\theta_e)/d\Omega_e$, first calculated by M.~Rosenbluth~\cite{Rosenbluth}, can be written as the product of the cross section for scattering from a structureless object and a structure-dependent term that depends on the Sachs magnetic and electric form factors~\cite{Sachs:1962zzc}, G$_{_M}($Q$^2)$ and G$_{_E}($Q$^2)$, which are, respectively, determined by the spatial distributions of magnetization and charge in the nucleon:
\begin{equation}
\frac{d\sigma(\theta_e)}{d\Omega_e} = \frac{d\sigma_{_{Mott}}}{d\Omega_e} \cdot \frac{\tau \, {\rm G}_{_M}^2({\rm Q}^2) + \varepsilon \, {\rm G}_{_E}^2({\rm Q}^2)}{ \varepsilon (1+\tau)}\ \ 
\end{equation}
Here ${d\sigma_{_{Mott}}}/{d\Omega_e}$ is the Mott cross section for the scattering of an electron with incident energy $E_e$ from a spinless structureless object of mass $M$ into a solid angle $d\Omega_e$~\cite{mott}, $\theta_e$ is the scattering angle of the electron, $Q^2 = 4 E_e \cdot E_e' \sin^2(\theta_e/2)$ is the negative four-momentum transfer squared, $E'_e$ is the energy of the scattered electron,  $\varepsilon \equiv 1/\left[1 + 2(1+\tau) \tan^2(\theta_e/2)\right]$ is the virtual photon polarization parameter, and $\tau \equiv Q^2/4M^2$.
The structure-dependent term is isolated in the reduced cross section,
$\sigma_{_R} \,=\, \tau \ {\rm G}_{_M}^2({\rm Q}^2) 
+ \varepsilon \ {\rm G}_{_E}^2({\rm Q}^2)$.
The traditional approach for disentangling the individual form factors, known as a ``Rosenbluth separation", is to make measurements at fixed \qsq~but different values of $\varepsilon$, corresponding to different electron beam energies and scattering angles. At high $Q^2$, however, the scattering cross section is dominated by $G_M$, which is multiplied by a factor of $\tau$, and it becomes difficult to isolate the contribution from $G_E$. 

An alternative to using a Rosenbluth separation to determine the elastic nucleon form factors is the so-called double-polarization method which measures the ratio $G_E/G_M$~\cite{Akhiezer, Arnold}. As long as the magnetic form factor $G_M$ is known, the double-polarization method provides an effective way to determine $G_E$. One variant involves a polarized electron beam and a polarized target.  Another variant includes a polarized electron beam, an unpolarized target and a ``recoil polarimeter" that measures the polarization of the recoil nucleon. 
An advantage of using a polarized target to measure $G_E/G_M$ is that the figure-of-merit does not decrease with increasing $Q^2$. In contrast, the analyzing power and the resulting figure-of-merit of a recoil polarimeter  decreases with increasing $Q^2$. 
For the case of the neutron, when determining $G_E^n$, there is a huge advantage to using a polarized target because polarized gaseous $^3$He targets can tolerate relatively high beam currents. For the case of the proton, however, when determining $G_E^p$ below 15~\gevsq, the use of a recoil polarimetry is more advantageous because polarized proton targets based on dynamic nuclear polarization of NH$_3$ can  only tolerate beam currents on the order of $100\rm\,nA$.

\subsection{Proton \gep/\gmp~ratio measurements with SBS apparatus}
\label{sec:gep}

The first experiment for which the SBS was designed, the SBS GEp experiment, was a measurement of the ratio of the electric and magnetic form factors of the proton, $G_E^p/G_M^p$, and was proposed in 2007~\cite{GEp}.  
The experiment used the double polarization method and was based on a longitudinally polarized electron beam incident on an unpolarized liquid hydrogen target along with a measurement of the polarization of the recoil proton. The configuration used in the hadron arm was discussed in Section~\ref{sec:sbs} and shown in Fig.~\ref{fig:SBS-GEp}.  
The electron arm consisted of ECal, which was described in Section~\ref{sec:ecal} .

The polarization of the recoil proton has two components: one parallel to the proton momentum, $P_\parallel$, and another that is oriented in the scattering plane and transverse to the proton momentum, $P_\perp$. The ratio of these two polarization components is related to the ratio $G_E^p/G_M^p$:
\begin{equation}
 {\frac{G_E^p}{G_M^p}} = - {\frac{P_\perp}{P_\parallel}} {\frac{ E_e + E_{e'}}{2 M_p}} \tan {\frac{\theta_e}{2}} \ \ .
\label{eq:gepasym}
\end{equation}
The quantities $P_\perp$ and $P_\parallel$  are determined using a ``recoil polarimeter" that is based on scattering the recoil proton  from an (unpolarized) polyethylene analyzer.

The recoil polarimeter works on the principle that, for protons polarized transverse to their momentum, an azimuthal asymmetry results from a spin-orbit interaction with the nuclei in the analyzer material. 
After scattering from the hydrogen target, the proton passes through the SBS magnet with its field oriented close to the scattering plane. 
This causes the spin to precess resulting in a vertical spin component proportional to $P_\parallel$. 
The component $P_\perp$ is largely unaffected.
As the proton scatters from the analyzer, due to the spin-orbit interaction, there is an up/down asymmetry proportional to $P_\perp$ and a left/right asymmetry proportional to $P_\parallel$.
While there are multiple effects (particularly instrumental) that can affect the azimuthal distribution after the analyzer, the asymmetry of interest is easily isolated by rapidly flipping the helicity of the incident electrons at 30~Hz. 
Another critical aspect of the technique is that the value of the analyzing power and the degree of the electron-beam polarization both cancel in the determination of the ratio $P_\perp/P_\parallel$ of Eq.~\ref{eq:gepasym} and the form factor ratio ${G_E^p}/{G_M^p}$.

\begin{figure}[!htb]
\begin{center}
\includegraphics[trim = 0mm 0mm 0mm 0mm, angle=0, width=0.75\textwidth]{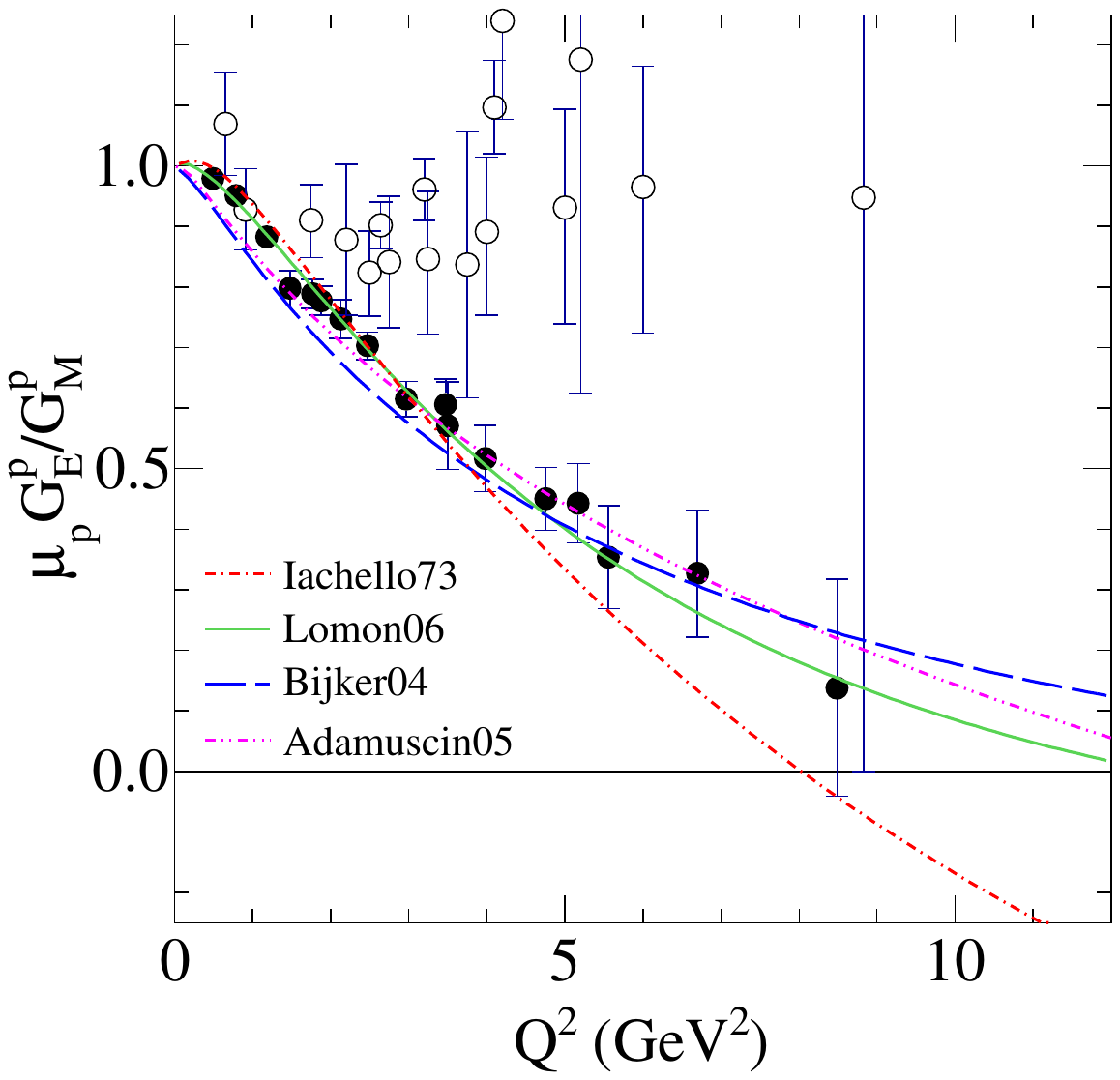}
\caption{The results on \gep/\gmp~from polarization transfer experiments. Figure is from ref.~\cite{GEp-all}. }
\label{fig:GEp/GMp}
\end{center}
\end{figure}

The underlying experimental technique used in the SBS measurement of $G_E^p/G_M^p$ was the same as that used by Jones {\it et al.} in ref.~\cite{Jones} when they discovered that the ratio $G_E^p/G_M^p$ decreases sharply with increasing $Q^2$ rather than remaining constant as had been expected previously. 
The result was confirmed in several additional polarization transfer experiments~\cite{GEp-all} as shown in Fig.~\ref{fig:GEp/GMp}.  
The marked discrepancy with results from Rosenbluth determinations is now believed to be largely due to a significant contribution to the scattering amplitude from the two-photon exchange~\cite{TPE}.

The SBS GEp experiment recently completed data taking and will determine the ratio $G_E^p/G_M^p$ at \qsq~=~11~\gevsq\  (see the updated proposal in ref.~\cite{GEp-update}).

\subsection{Neutron magnetic form factor} 

\begin{figure}[!htb]
\begin{center}
\includegraphics[trim = 0mm 0mm 0mm 0mm, angle=0, width=0.5\textwidth]{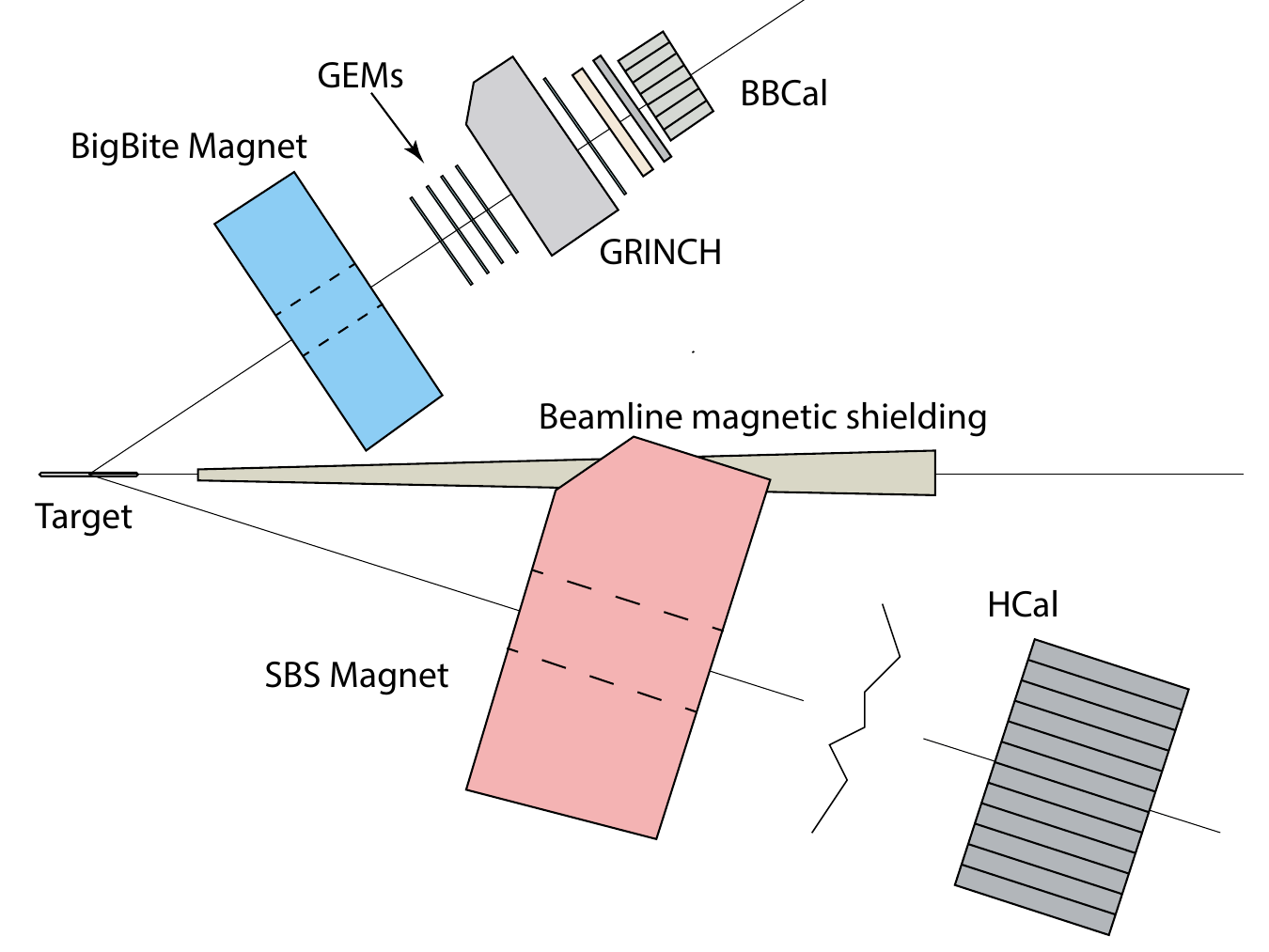}
\caption{The typical layout of a high \qsq~elastic form factor experiment for \gmn/\gmp~and \gen/\gmn.}
\label{fig:layout}
\end{center}
\end{figure}

As discussed earlier, the components of SBS can be reconfigured in combination with other existing equipment to perform a variety of experiments.  
One such experiment was a measurement of $G_M^n$ to high values of $Q^2$~\cite{GMn-1}. 
A schematic representation of the layout used for the SBS $G_M^n$ experiment (as well as GEn-II) is shown in Fig.~\ref{fig:layout}.  
The electron was detected using the BigBite spectrometer as described in Section~\ref{sec:bigbite} (see also ref.~\cite{BBCal}).
The hadrons, in this case both protons and neutrons, were detected in coincidence using HCal, which was described in Section~\ref{sec:hcal}. 
The SBS magnet was placed close to the target and deflected protons relative to neutrons, making particle identification relatively straightforward.

The SBS experiment to measure $G_M^n$  used the “ratio method”~\cite{Durand}, probably the most productive  of a variety of methods that have been used to measure $G_M^n$ over a wide range of $Q^2$~\cite{GMn-DESY, GMn-SLAC, GMn-Xu, GMn-Bates, GMn-CLAS}. 
The ratio method relies on the detection, using the same apparatus, of both  protons and neutrons that recoil from a liquid deuterium target. 
There are substantial corrections that need to be applied when trying extract the elastic cross sections for protons and neutrons using quasi-elastic scattering from deuterium.
These corrections are nearly identical, however, for both protons and neutrons and thus largely cancel in forming the ratio. 
Furthermore, by forming a coincidence between the electron and the quasi-elastically scattered nucleon, background is kept to a manageable level.  

The SBS $G_M^n$ experiment benefits from multiple factors when compared to other non-SBS based experiments.  
As already noted in the introduction and as is illustrated in Fig.~\ref{fig:Landscape}, the combination of luminosity and solid angle ensures that the statistics from an SBS-based experiment will be be competitive with any other spectrometer system at JLab.  
This is even true when one takes into account that the SBS-based $G_M^n$ experiment essentially measures one $Q^2$ point at a time. 
As noted above, the SBS magnet enables very effective particle ID, and HCal helps to reduce uncertainty in the neutron/proton detection efficiency.  
It is also worth noting that the simplicity of the SBS layout also helps in ensuring that the fiducial detection volumes for protons neutrons are well understood and effectively the same.  

At the time of this writing the analysis of the SBS $G_M^n$  experiment is nearing completion, and the projected errors are likely to be in the range of 2-3\%.  We also note that, with the success of the SBS $G_M^n$ run, there is every reason to be optimistic that a future experiment could extend the $Q^2$~range with the same apparatus and comparable accuracy to $18\rm$\gevsq~with a 3-4 week run, as has been proposed~\cite{GMn-2}.

\subsection{Neutron \gen/\gmn~ratio}

A second experiment that used the basic layout shown in Fig.~\ref{fig:layout} was the primary SBS experiment to measure the ratio of the electric and magnetic form factors of the neutron, $G_E^n/G_M^n$, an experiment we will refer to herein as GEn-II. 
Like the SBS GEp experiment, GEn-II used the double-polarization method, only in this case, both a polarized beam and a polarized ($^3$He) target was used and the  polarization of the recoil neutron was not measured. 
The principle difference in the configurations of the SBS $G_M^n$ experiment and GEn-II is that a polarized $^3$He target was used instead of a liquid deuterium target. GEn-II nearly tripled the range of \qsq~over which that ratio  $G_E^n/G_M^n$ has been accurately measured. This dramatic increase is particularly important because it brings the range over which $G_E^n/G_M^n$~is known into the same basic range over which $G_E^p/G_M^p$~is known. 
As will be discussed in the next subsection, an important reason to study the elastic form factors of the neutron and proton over similar ranges of $Q^2$ is that it makes it possible to perform a flavor decomposition of the u- and d-quark contributions to the elastic form factors.  

In GEn-II, polarized electrons were scattered from a polarized $^3$He target and the scattered electrons and recoil neutrons were measured in coincidence. 
The hadron arm included the SBS dipole magnet at 2 meters from the target and facilitated distinguishing neutrons from protons by deflecting the protons downward while the neutrons traveled straight.  
The neutrons were detected by the hadron calorimeter, HCal, which was situated roughly 17~meters from the target. 
The SBS dipole also significantly reduced the background reaching HCal by deflecting lower-energy charged particles. 
The electron arm used the BigBite spectrometer and provided momentum resolution of 1--2\% depending on the kinematic setting. 

For the idealized case of elastic scattering of 100\% polarized electrons off 
of free 100\% polarized neutrons, a target polarization, $\vec P_{\mathrm He}$, that is oriented at $90^{\circ}$ with respect to the 3-vector momentum transfer $\vec q$, $G_E^n$~is related to the double spin asymmetry $A_{en}$~\cite{Donnelly} by \\
\vspace{-1. cm}
\begin{center}
\begin{equation}
A_{en} = \frac
{-2\sqrt{\tau(\tau+1)} \tan(\theta_e/2) \,\cos\phi^*  \,G_E^n/G_M^n} 
{\left ( G_E^n/G_M^n \right ) ^{^2} \,+\, \tau
\left [ 1 \,+\, 2(1+\tau)\tan^2(\theta_e/2) \right ] } 
\end{equation}
\end{center}
where $\theta_e$ is the scattering angle of the electron,  $\tau = Q^2/4M^2$, 
and $\phi^*$ is the angle between the electron scattering plane and the ($\vec P_{\mathrm He}, \, \vec q$) plane.      

At the time of this writing, data taking for GEn-II is complete and the analysis is well underway.  Data were taken for three values of $Q^2$, $3.0\rm$\gevsq, $6.8\rm$\gevsq and $9.8\rm$\gevsq~and will be valuable both in their own right as well as for the flavor decomposition described in the next subsection. 

\subsection{Flavor decomposition of the nucleon form factors}

We suggested earlier that understanding non-perturbative QCD requires, at some level, unveiling the roles of the quarks and gluons in nucleon structure.  As we describe below, having accurate measures of all four elastic nucleon form factors makes it possible to determine the contributions to those form factors from individual quark flavors, something that is clearly a step in the right direction. 

In the one-photon exchange approximation, the amplitude for electron-nucleon elastic scattering can be written $ {M^{\rm^{EM}} = -(4\pi\alpha/Q^2)l^{\mu}\,J_{\mu}^{\rm^{EM}}}$, where $\alpha$ is the fine structure constant,  
$l^{\mu}= {\overline{e}} \gamma^{\mu} e$ is the leptonic vector current, and 
\begin{equation}
J_{\mu}^{\rm^{EM}} = \langle {p (n) \,|{\textstyle{+\frac{2}{3}}}{\overline{u}}\gamma_{\mu}u 
+ {\textstyle{\frac{-1}{3}}}{\overline{d}}\gamma_{\mu}d|\, p (n)} \rangle
\label{eq:jmu}
\end{equation}
is the hadronic matrix element of the electromagnetic current operators for the proton (neutron).
Here we have neglected to include a term in Eq.~\ref{eq:jmu} for strange quarks, a possibility that is the subject of both a fully approved future SBS-based experiment~\cite{sFF} as well as another possible experiment that is under development and discussed in Section~\ref{sec:pves_strange}. 
From Eq.~\ref{eq:jmu}, and with the additional assumption of charge symmetry, it follows that the contribution to the Dirac and Pauli form factors of the nucleon from u- and d-quarks is given by
\begin{center}
$F_{1(2)}^u \, = \, 2\,F_{1(2)}^p \,+\, F_{1(2)}^n$
  \hskip .05in and \hskip .05in 
$F_{1(2)}^d \, = \, 2\,F_{1(2)}^n \,+\, F_{1(2)}^p$, \\
\end{center} 
where $F_{1(2)}^n$ and $F_{1(2)}^p$ are the Dirac (Pauli) form factors of the neutron and proton, respectively. 
Here we follow the usual convention that $F_{1(2)}^u$ and $F_{1(2)}^d$ represent the u- and d-quark contributions in the proton.

With the determination of the electric form factor of the neutron, $G_E^n$, to 3.4~\gevsq~\cite{GEn-1},  it became possible, for the first time, to perform a flavor decomposition of the elastic nucleon form factors to values of \qsq~well above 1~\gevsq. 
This decomposition showed that scaling with \qsq~was quite different for up quarks and down quarks. 
This behavior has been interpreted as evidence supporting the importance of diquark correlations in nucleon structure \cite{Barabanov}.  
The flavor separated form factors also play a key role in constraining GPD models, and thus impact some of the most important questions in understanding nucleon structure.

A central motivation driving the SBS physics program is to ensure that all four of the elastic nucleon form factors are known to as high a value of \qsq~as is practical.  
The magnetic form factor of the proton, $G_M^p$, has historically been the most studied and was also the subject of a recent experiment at JLab up to \qsq=15.75~\gevsq~\cite{GMp12}.  
The ratio of the electric and magnetic form factors of the proton, $G_E^p/G_M^p$, as discussed earlier, has been the subject of multiple experiments, and indeed the discovery of the nearly linear drop in $G_E^p/G_M^p$ with \qsq~\cite{Jones} played a key role in recent interest in elastic nucleon form factors.  
The SBS physics program will extend accurate measurements of $G_E^n/G_M^n$~to nearly 10~\gevsq, and will improve the accuracy with which both $G_M^n$ and $G_E^p/G_M^p$ are known at high $Q^2$. 
This coherent approach will make it possible to determine the flavor separated elastic form factors to roughly 10~\gevsq. 
The $Q^2$ behavior of the flavor separated form factors up to 3.4~\gevsq~has already had a significant impact on our physical understanding of nucleon structure. 
Understanding how the flavor separated form factors behave at significantly higher $Q^2$ is clearly of great interest.

\section{Transversity distribution functions in the polarized neutron}

We describe next an experiment to measure single-spin asymmetries in semi-inclusive deep inelastic scattering (SIDIS), measurements that are sensitive to transverse momentum dependent (TMD) distribution functions and phenomena such as the Sivers and Collins effects and the transverse polarization of quarks in a transversely polarized target~\cite{Sivers, Collins}.
The SBS SIDIS experiment, as we will refer to it, has the highest scientific ranking (from the JLab Program Advisory Committee) of any of the planned SBS experiments that, at the time of this writing, have not yet run.

The polarized SIDIS process, $\vec N(e,e'h)X$, in which a single hadron is observed in coincidence with the scattered electron in a DIS collision, was investigated by HERMES~\cite{HERMES} and more recently by COMPASS~\cite{COMPASS}. 
It is a powerful method for the multi-dimensional imaging of quark motion in the nucleon. 
The electron undergoes a hard scattering with a quasi-free quark in the target nucleon, which then ``fragments'' into color-neutral hadrons as it recoils from the hard collision. 
Of particular interest is the ``leading hadron", here a pion or kaon, with an energy $E_h$ that represents a large fraction, $z$, of the energy available from the collision where $z=E_h/\nu$ and $\nu$ is the usual energy loss of the electron. 
In the kinematic regime in which $P^h_{\perp}$ (the component of the leading hadron's momentum that is transverse to the virtual photon direction) is small compared to the magnitude of $Q$, the cross section can be expressed in terms of transverse-momentum-dependent (TMD) distribution functions and fragmentation functions. 
When the target nucleon is transversely polarized, there is rich physics to be learned by looking at the azimuthal dependence of the lead hadrons around the virtual photon direction including the transverse polarization of the nucleon's constituent quarks and the now well-known Sivers effect~\cite{Sivers}.

The SBS SIDIS experiment~\cite{SIDIS} will provide unprecedented sensitivity to TMD-related physics for reasons related to our discussion of Fig.~\ref{fig:Landscape}.  
There is strong kinematic focusing of the high-$z$ leading hadrons along the direction of the virtual photon. 
Thus, for a given value of \qsq, with SBS centered on the momentum transfer $\vec q$, the medium-sized acceptance is more than adequate to capture the vast majority of events of interest.  Furthermore, the large vertical acceptance of SBS and HCal together with the flexibility of setting the target direction guarantees that the azimuthal coverage will be excellent. 
When combined with the high luminosities that are possible, up to nearly $\rm10^{38}\,cm^{-2}s^{-1}$ with a polarized $^3$He target, SBS SIDIS will have statistical power that is 10-100 times larger than much of the existing data, particularly in the high-$x$ range that is particularly accessible at JLab.

\section{Single pion photo-production and wide angle Compton scattering}
\label{sec:wac}

\subsection{Single pion photo-production}

Meson photo-production from the nucleon has been the subject of physics interest since the 1950s.
Many experiments have been performed at various electron beam facilities to study exclusive photo-production reactions.
For the processes $\gamma p \rightarrow \pi^{+} n$, $\gamma n \rightarrow \pi^{-} p$, and $\gamma p \rightarrow \pi^{0} p$, perturbative QCD predicts~\cite{pQCD} that $d\sigma/dt \propto s^{-7}$. 
Although this model is a fairly good representation of the  scaling that is observed experimentally, it fails dramatically in explaining the absolute cross sections.
A leading-twist result from a GPD-based calculation~\cite{Pion-2000} also underestimates the observed cross section by multiple orders of magnitude (see ref.~\cite{Pion-CLAS}).
A newer GPD-based calculation~\cite{Pion-2021}, which takes into account twist-3 effects, was able to reproduce the cross section and makes a novel prediction for the double polarization observables: \ALL = -\KLL. 
Confirmation of this prediction would be powerful evidence supporting the recent GPD-based description of pion photo-production.

The SBS physics program includes two approved experiments to study single pion photo-production.
One experiment, that has already run, measured the double-polarization asymmetry $K_{LL}$~\cite{Pion_KLL} using  a polarized electron beam, an unpolarized liquid deuterium target and a recoil polarimeter.  
The other, that will presumably run in the next group of SBS experiments, will measured the double polarization asymmetry $A_{LL}$~\cite{Pion_ALL} using a polarized electron beam scattering off a polarized $^3$He target.

\begin{figure}[!htb]
\begin{center}
\includegraphics[trim = 0mm 0mm 0mm 0mm, angle=0, width=0.5\textwidth]{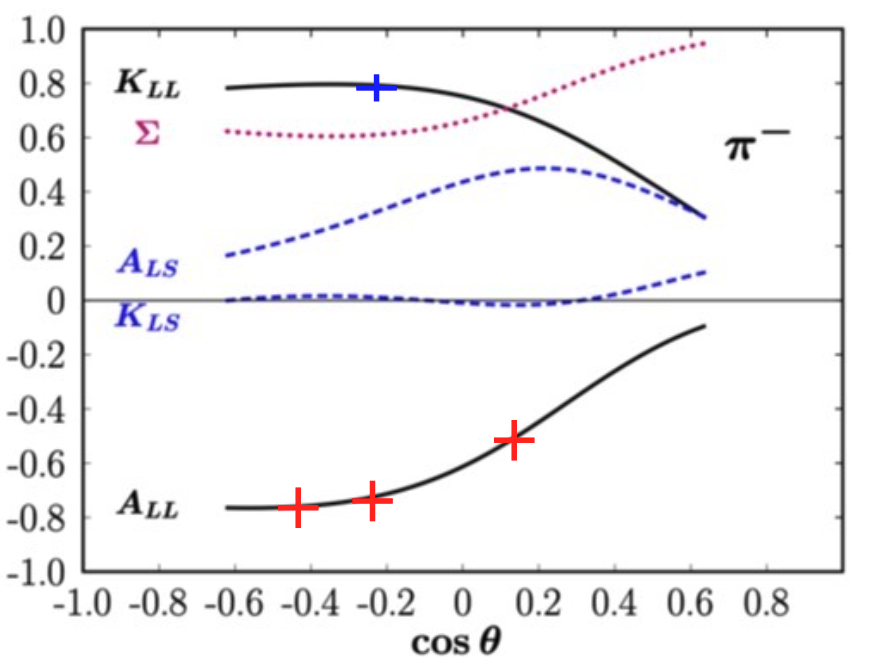}
\caption{The projected data points from the experiments~\cite{Pion_ALL, Pion_KLL}.}
\label{fig:pion}
\end{center}
\end{figure}

\subsection{Wide angle Compton scattering}
\label{sec:wacs}

Real photon Compton scattering from the nucleon with $s$, $-t$, and $-u$ values large compared to $\Lambda_{_{QCD}}^2$, Wide Angle Compton Scattering (WACS), is a hard exclusive process which provides access to information about nucleon structure complementary to high \qsq~elastic form factors and Deeply Virtual Compton Scattering.  
The first WACS experiment from the proton was made at Cornell~\cite{Cornell}.
Much more accurate results obtained at JLab~\cite{WACS-cross,WACS-KLL-1,WACS-KLL-2} are in unambiguous agreement with the handbag mechanism~\cite{DK-2013, WACS-PK}.

There are currently two fully approved WACS experiments at JLab that, with the 11-GeV beam, will be able to access higher values of $s$~\cite{WACS-ALL, WACS-3}. 
One experiment, based on the Compact Photon Source described in Section~\ref{sec:CPS} and the BigBite spectrometer (with the SBS hadron calorimeter) as a proton arm, will measure the double polarization asymmetry with a polarized proton target with both $s$ and $t$ in the range of reliable GPD-based predictions.
The other experiment is a measurement of the cross section with $s$ up to 20~\gevsq~using the Hall C HMS
spectrometer as a proton arm and a high resolution calorimeter as a photon arm~\cite{NPS}.
With the SBS apparatus as a proton arm there is a possibility to measure the WACS cross section from the neutron, which has never been investigated.
The GPD-based calculation of the neutron WACS, published in ref.~\cite{WACS-PK}, predicts a large reduction of the cross section relative to the proton case.

===================================
\section{Potential experiments with a Double Open SBS}
\label{sec:double}

There are currently several new experiments under consideration that would use the DOSBS: high \qsq~$\phi$-meson electro production, DVCS with polarimeters,  the charged weak $p\rightarrow\Delta^0$ transition form factor and parity violation at high \qsq~for elastic and deep inelastic scattering.

\subsection{Nucleon mass (gluon) distribution}

Recent advances in our understanding of both Generalized Parton Distributions (GPDs) and holographic QCD has made it possible to interpret experimental observations in terms of the so-called gravitational form factors which are the matrix elements of the proton's QCD Energy Momentum Tensor (EMT). These advances have resulted in determinations of quantities such as the pressure distribution inside a nucleon~\cite{burkert} and more recently, the distribution of the nucleon's mass~\cite{gff}. The results presented in ref.~\cite{gff}, involved studies of the photoproduction of the $J/\psi$ near threshold. It appears likely that studies of the electroproduction of the $\phi$ meson near threshold would be similarly productive, and has been proposed as a letter of intent using the SBS spectrometer~\cite{deepPhi}. This experimental effort, which faces statistical challenges, is a good example of an experiment that would strongly benefit from the use of the DOSBS.  
The proposed experiment~\cite{deepPhi} expects to reach \qsq~=~8~\gevsq.

\subsection{Charged weak current transition form factor}

Recently, a Letter-of-Intent~\cite{tAVFF} was submitted to JLab PAC54 that would use the DOSBS to measure the cross section for electrons scattering from protons through a charged weak current, that is, through the exchange of a $W^-$.  Thus, the reaction of interest and the subsequent decay can be written as:
 $ \,\,\, \vec e + p \rightarrow \nu_e + \Delta^0 \rightarrow \nu_e + p + \pi^- \, $.
The decay products of the $\Delta^0$ are detected and used to reconstruct the energy of the beam as a means of isolating the charged current interaction. Here the large acceptance of DOSBS is critical for the efficient detection of the proton and the $\pi^-$.  While there will be a large background, particularly from the production of two pions, the scattering due to the charged weak current is parity violating making it much more straight forward to distinguish the reaction of interest~\cite{tAVFF}.
The experiment proposes to measure of range of momentum transfer to the proton of 1-2~\gevsq.

An important motivation for this experiment is a pressing need within the neutrino physics community to better constrain neutrino-proton cross sections over a wide range of momentum transfer.  The reaction can be understood theoretically within the framework of GPDs, and the experiment will provide valuable constraints on those GPDs. Finally, the experiment provides an elegant way to explore neutrino-proton interactions without needing to construct a massive neutrino detector.

\subsection{Parity violation in elastic electron scattering}
\label{sec:pves_strange}

About 35 years ago, parity violation in elastic electron-proton scattering was proposed as a technique for determining the strange-quark contribution to the elastic form factors~\cite{BM-1989,DB-1989}. 
Multiple experiments followed that measured parity violation (PV) in elastic $e-p$ scattering at relatively small momentum transfer that established strong upper limits on the strange-quark effects~\cite{SAMPLE,G0-1, G0-2,MAMI-PV,Happex-1, Happex-2, Happex-3}. 
It can be argued, however, that strange-quark effects could be significantly higher at larger momentum transfer where no experimental limits exist.  
Recently, an experiment~\cite{sFF} was approved to search for strange-quark contributions to elastic scattering at 2.5~\gevsq\ based on the detection of a scattered electron  and the recoiling proton in coincidence (using SBS detector components); the electron arm in that experiment has a solid angle of 30-40~msr. 
The very large solid angle (0.26~sr) and good momentum resolution (0.5\%) that is projected for DOSBS would make it possible to reach even higher \qsq~in a single-arm mode detecting scattered electrons. 
The combination of the two experiments at 2.5~\gevsq~would make possible an LT separation of the strange form factors.

\subsection{Parity violation in deep inelastic electron scattering at high momentum transfer}
An interesting possibility is the use of DOSBS for the study of parity violation in deep inelastic scattering (PVDIS)~\cite{PVDIS}.

Parity violation in electron scattering was first discovered in deep inelastic scattering and played a pivotal role is establishing what became the Standard Model~\cite{Prescott-1978}. 
Those first pioneering experiments, conducted at SLAC, were primarily sensitive to the vector coupling of quarks to the weak neutral current.  PVDIS was not studied again for nearly 40 years until an experiment at JLab observed for the first time evidence for a non-zero axial-vector coupling of quarks to the weak neutral current~\cite{PVDIS-3}. In a certain sense, it was the first experiment to show that quarks have a "handedness". The combined solid-angle acceptance of the two high-resolution spectrometers used in ref.~\cite{PVDIS-3} was roughly 12~msr.

An experiment to study of PVDIS using DOSBS could have a solid-angle acceptance of around 260-msr solid, roughly 20 times that described in ref.~\cite{PVDIS-3}, and would also have a much wider momentum acceptance.
Such an experiment could use the detector package currently under consideration for DOSBS along with transition radiation counters for electron identification~\cite{TRD}. 
As has been discussed at some length in connection to the proposed SoLID spectrometer, there is a rich variety of physics that can be explored using PVDIS~\cite{SoLID-PVDIS}. For example, PVDIS at high Bjorken $x$ is sensitive to physics beyond the Standard Model at very high energy scales and is largely insensitive to hadronic structure.  In related studies, PVDIS can be used to study the important quantity $d/u$ with very little influence from higher-twist effects.  Studies of charge symmetry violation is another possibility. The potential of DOSBS for PVDIS studies is clearly worth further examination.

\section{Conclusion}

This paper presents a broad program of investigations that are possible using the SBS, its associated detector systems and the JLab accelerator beam. We discuss multiple examples in which, at high momentum transfer, there are strong correlations between the scattered electron and the high-momentum recoiling nucleon (or a high-z meson) leading to an excellent figure-of-merit even with modest solid angles in each detector arm. In many cases, because of the possibility of using high luminosities, the advantages of even larger solid angles are less clear cut. The paper also presents key ideas underlying the instrumentation developed for SBS and, through multiple examples, we underscore the flexibility of the SBS and associated systems.

We give brief summaries of the experiments that were completed during the several years that comprised the first set of SBS experiments as well as descriptions of the the fully approved and often high priority experiments that are to come.  Finally, we explore briefly some of the ideas that are under development for SBS that are still early either in their conceptualization or the approval process.

\acknowledgments{We appreciate the contributions given to the SBS physics program by our colleagues of the SBS coordination committee and by the members of the SBS collaboration.
This material is based upon work supported by the U.S. Department of Energy, Office of Science, Office of Nuclear Physics under Contract No. 89243126CSC000213 (B.W.), DE-FG02-01ER41168 (G.C.).

\title{References}

\bibliography{main.bib}

\end{document}